\documentclass[a4paper,12pt]{article}
\usepackage{epsfig}
\usepackage{amssymb}
\begin{document}
\thispagestyle{empty}

%____________________________________________
%\newcommand{\JP}    {jet probability}
\newcommand{\etal}  {{\it{et al.}}}  % must have all the {}
\def\Journal#1#2#3#4{{#1} {\bf #2}, #3 (#4)}
\def\PRD{Phys.\ Rev.\ D}
\def\NIMA{Nucl.\ Instrum.\ Methods A}
\def\PRL{Phys.\ Rev.\ Lett.\ }
\def\PLB{Phys.\ Lett.\ B}
\def\EPJ{Eur.\ Phys.\ J}
\def\IEEETNS{IEEE Trans.\ Nucl.\ Sci.\ }
\def\CPCD{Comput.\ Phys.\ Commun.\ }
%____________________________________________

%\hfill {\LARGE\bf DRAFT}
%\smallskip

%\hfill {\large\bf \today}
\bigskip
%\bigskip

%{\LARGE\bf
{\Large\bf
\begin{center}
Intuitive  study to the scalar boson stars formation
\end{center}
}
\vspace{0.1 cm}
%\vspace*{\fill}

\begin{center}
{ G.A. Kozlov  }
\end{center}
%\vspace{1 cm}
%\vspace*{\fill}
%\date
\begin{center}
\noindent
 { Bogolyubov Laboratory of Theoretical Physics\\
 Joint Institute for Nuclear Research,\\
 Joliot Curie st., 6, Dubna, Moscow region, 141980 Russia  }
\end{center}
%\vspace*{\fill}
\vspace{0.1 cm}

 %\section*{Abstract}
 \begin{abstract}
 \noindent
 {We study the  feebly interacting dark matter with the Standard Model fields, where the preceding and the latter ones are the constituents of the bosonic cosmological objects, the scalar boson stars (BS). 
The scalar dark matter (SDM) massive fields can form the macroscopic scalar bound state when minimally coupled to gravity (GR). The star may have approximately a stationary form with an asymptotically flat geometry, and the maximal mass  of the BS is larger than the solar mass. 
We consider an effective description of a more complete model with the (thermo)dynamic potential of the BS at finite temperature. 
%for an arbitrary number of dark matter spinor particles with arbitrary masses and chemical potentials. 
The formation of the BS may also be explained in terms of the electric fluxes of the hidden vector field under the influence of the SDM field as the cosmological dynamical quantity minimally coupled to GR. 
The SDM fields may fluctuate with the temperature and is established around the equilibrium state at some weak background scale.    
 }

%\vspace*{\fill}

\end {abstract}

%\vspace*{\fill}

%\newpage
%\tableofcontents

%\newpage
%\section{Introduction}

\bigskip

\section{Introduction}
The  feebly interacting dark matter (DM) at high temperature $T$ and the energy density $\rho$, corresponding to freeze-out point of the matter state, is one of the close focuses to the researchers in DM physics and cosmology. It is assumed that the early Universe was occupied by free particles, like gluons, quarks as well as the DM species with different spins. As the $T$ was cooled, there were experienced the cosmological critical point (CP) and the phase transitions  associated with the symmetry breaking and with the separation of the unified fundamental interactions from one another.

In the paper we follow the speculations  concerning the potentially important role played by the evolving scalar DM fields and
the interactions between these fields with the Standard Model (SM) fields being the constituents of the bosonic cosmological objects called the boson stars 
(see, e.g.,  [1] and the refs. therein). These stars may carry the spin 0, the scalar boson stars (BS) [2,3], or the spin 1, the Proca stars [4]. The axion boson stars with the ultra-light bosons of the masses $\sim O(10^{-10} eV)$ have been considered in many papers (see, e.g., [5] and the refs. therein). The BS may be understand as a stationary configuration of the DM and the SM scalar fields bounded by gravity. The stability of the BS with the lifetime of the order $\sim O(10^{-7} s)$  for the long-lived particles (LLP) [6], can arise naturally due to gravitational collapse (of the scalar fields) in the final end-state of the star. 
The scalar dark matter (SDM) may also interact with the spin 1/2 DM $\chi$ - particles which are treated as a stable ones because of the  $\mathbb{Z}_2$ symmetry.  
The CP is followed at $T\sim $ MeV - GeV range with a few tens of $\mu$sec after the big-bang, in which the SDM sector and the Higgs sector were confined into the BS and the scale symmetry was broken. Although we can only suppose what happed in the CP phase, there has been no observational signature until recently showing the physical processes (inside the BS) operated at this epoch. Even though various strong indirect evidences of the BS have been established through astrophysical and cosmological observations, the absence of direct detection leaves the properties of the fields inside and the mechanism of the BS formation still largely unknown.

The one of the candidates to DM is the SDM. Since these scalar fields are created in the phase with the cosmological CP before the recombination epoch, they are almost free from the gravitational dissipation and can even distribute over the galactic halo as well, making a contribution to the total energy budget of the Universe $\Omega_{univ}$ if they survive until our days. The stage after the CP if happed is characterised by the formation of the BS as well as the stellar black holes, fermion stars, neutron stars, etc. 
%At the epoch of cosmological CP, there is the background related with photons and light leptons, where the energy density of these particles is very huge,  $\rho_{\gamma l} \sim %T^4\sim O(10^{39} GeVcm^{-3})$ compared to that of the DM energy density $\rho_\odot \simeq 0.43 \,GeV\,cm^{-3}$ of the solar system [7] or $\rho_\odot \simeq 0.36\pm 0.02 %\,GeV\,cm^{-3}$ as the best-fit dark halo profile result [8]. 
The DM mass should be in the weak scale range $\sim 0.1 - 10$ TeV based on the observed DM abundance $\Omega_{DM} h^2\rightarrow \Omega_{Planck} h^2 \simeq 0.2$ [7] in the early Universe with the total thermal relic abundance density $\rho_\chi\sim \Omega_{DM} h^2\cdot \rho_c\sim O(10^{-6} GeV cm^{-3})$ where $\rho_c\sim 1.05 \cdot 10^{-5} GeV cm^{-3}$ is the critical density of the Universe [8]. Here, it is assumed that $\rho_\chi$ consists of the contributions from the decay of the SDM and from freeze-in production of $\chi$
%the DM annihilation rate has the sizeable values of the weak scale cross section 
where $\rho_\chi/\rho_\odot \sim O(10^{-7})$ with  $\rho_\odot \simeq 0.43 \,GeV\,cm^{-3}$ being the local DM energy density of the solar system [9] ( $\rho_\odot \simeq 0.36\pm 0.02 \,GeV\,cm^{-3}$ as the best-fit dark halo profile result [10]).

At temperatures and the DM densities below the phase transition point, the BS is characterised by the phenomena of scale invariance breaking, the so-called "confinement" of scalar particles inside the BS,  and the stability of the star. This is due to Abelian Higgs-like effect of interaction between the SDM fields and the vector fields (e.g., photons, dark photons, "dark gluons") entered the evolving star. 
%In some sense, the SDM is the dynamical dark energy. 
%The scale symmetry is broken spontaneously by $\bar\chi\chi$ condensate, and also explicitly by both the mass $m$ of $\chi$ and the scale anomaly induced by the same $%\bar\chi\chi$ condensate. 
The scale invariance breaking is dictated by the relation 
%\begin{equation}
%\label{e6}
 $ \langle 0\vert D^\mu (0)\vert X_\star (q)\rangle = -i\Lambda_\star q ^\mu, $
%\end{equation}
where $D_\mu(x)$ is the dilatation current, $X_\star (q)$ is the operator associated with the real scalar field of the BS,
$\Lambda_\star \sim O(M_{Pl})$ is the decay constant of the BS which is of the order of the Planck mass $M_{Pl} \simeq 1.2\cdot 10^{19}$ GeV which is the only fundamental scale in Nature. The electroweak (EW) scale is the consequence  of $M_{Pl}$ because of the non-perturbative phenomena in gravity, where the SM Higgs mass is related to $M_{Pl}$ as $m_h = M_{Pl}\, e^{-a}$ with $ a\sim 40$.
The $D_\mu$ is not conserved and the following properties to the trace of the energy-momentum tensor $\Theta_\mu^\mu$ are in force
%\begin{equation}
%\label{e7}
  $$\partial_\mu D^\mu = \Theta_\mu ^\mu = m\bar\chi\chi + \frac{\beta (\alpha_D)}{4\,\alpha_D(\tilde\mu)} B_{\mu\nu}B^{\mu\nu}, $$
% \end{equation}
where $\beta (\alpha_D)$ is the renormalisation-group $\beta$-function, 
$B_{\mu\nu} = \partial_\mu B_\nu - \partial_\nu B_\mu$ is the Abelian field strength tensor to the vector field $B_\mu $, $\alpha_D (\tilde\mu)$ is the renormalised DM coupling constant defined at some scale $\tilde\mu$.
The BS mass $M_\star$,  being of the order of the solar mass or even bigger [1,11], is defined from 
%\begin{equation}
%\label{e8}
 $\langle 0\vert \partial_\mu D^\mu (0)\vert X_\star (q)\rangle = -\Lambda_\star M_\star ^2. $
%\end{equation}
%The scale symmetry is broken spontaneously by $\bar\chi\chi$ condensate, and also explicitly by both the mass $m$ of $\chi$ and the scale anomaly induced by the same $%\bar\chi\chi$ condensate.
The DM $\chi$ may annihilate into two SDM particles $\phi$, either via exchanging $\chi$ in the t-channel, or directly with a point-like interaction with production of two $\phi$ particles where the latter may constitute to the BS and contribute to formation of the star. The scale symmetry is broken spontaneously by $\bar\chi\chi$ condensate, and also explicitly by both the mass $m$ of $\chi$ and the scale anomaly induced by the same $\bar\chi\chi$ condensate.

The mechanism of the formation of the BS being asymptotically with the flat direction is governed by interaction forces between the hidden fields with the SM sector plus gravity. A huge number of electric flux tubes, as the vortex excitations of the electric flux, 
 initiated by the hidden vector sector (the mediators between $\bar\chi$ and $\chi$), may be the key components of the field configuration inside the star under the influence of the SDM fields squeezing the flux into a tube with the radius $r_s$.  On the other hand, the ends of the tubes may meat each other to form a circle with the radius $R\sim$ size of the BS ($R >>> r_s$).

If the BSs can form and exist in the Universe, they might also form the compound composite systems with the Proca-stars and the axion boson stars which in turn may be produced non-thermally in the early Universe and 
would be identified through the novel gravitational-wave sources.
Assuming the existence of the Proca-star with the w.f. of the form $V_\mu =\bar\chi\gamma_\mu\chi$ and simultaneously  the axion-like star with the w.f. $A$, having the pseudo-scalar structure identified through the interactions with photons and gluons [12], the following operator equality for the BS operator $X_\star$ would be as [13]
%\begin{equation}
%\label{e5}
$ X_\star^2 = V_\mu^2 + A^2 $,
%\end{equation}
which is the special case of Fierz equations [14] as well.  To date, no boson stars have been firmly discovered yet. More studies on the BS from different directions are necessary to specify the star states.

In the paper, we present an analytic and rather intuitive approach to the BS study.
The effective model to formation of the BS in terms of the fluxes of the gauge boson fields under the influence of the SDM are considered in Sec.2 and in Sec.4 (with the gravity). In the Sec.3, the (thermo)dynamics of the BS with the Higgs portal is presented. The maximal mass of the BS, $M_\star^{max}$, and the lower bound on the critical temperature depending on  $M_\star^{max}$  are estimated. The paper is concluded in Sec.5.

\section{Formation of the BS}
Today not so much known concerning the mechanism of the BS formation, in particular, the binding forces between the fields inside the BS.
Rather than binding of the SDM (and the SM) particles, we want to understand what one might call the formation of the BS at some extreme conditions.  At high $T$ and $\rho$, where non-vanishing chemical potentials $\mu$ are assumed, the SDM becomes feebly coupled and at the vicinity of the CP these particles are no more interact, the scale symmetry is restored. One of the approaches to study the DM at extreme conditions can be realised through the scheme with topological defects which are absent in the SM, but are generic
in some effective models [15-17]. These defects exist only in the phase with spontaneously broken symmetry where the average expectations of the SDM fields are non-zero. 
%The minimal model known where the topological defects arise is the Abelian Higgs-like model [14]. For example, the topology of the rank-$N$ $SU(N)$ manifold and that of its Abelian %subgroup  $[U(1)]^{N-1}$ are different, and since any gauge transformations are singular, one might introduce the string-like object by performing the singular gauge transformation %with an Abelian gauge field $B_\mu$ [15],
%$B_\mu (x)\rightarrow B_\mu (x) + (g/4\pi) \partial_\mu \omega (x)$. Here,  $\omega (x)$ is the angle subtended by the closed space-time curve ("string") at any point $x = (x^0, x^1)$, %and $g$ is the constant 
%dual to the gauge coupling constant $e$, $g =2\pi/e$, 
%being responsible for the flux inside the "string". The Abelian field strength tensor $B_{\mu\nu}$ transforms as $B_{\mu\nu} (x) \rightarrow B_{\mu\nu}(x) + \tilde G_{\mu\nu} (x)$, where %$\tilde G_{\mu\nu} (x) = (g/4\pi) [\partial_\mu ,\partial_\nu] \omega(x)$ is nothing other but
The singular gauge transformation where the $B_{\mu\nu}$ transforms as $B_{\mu\nu} (x) \rightarrow B_{\mu\nu}(x) + \tilde G_{\mu\nu} (x)$
%with the field $B_\mu$, $B_\mu (x)\rightarrow B_\mu (x) + (g/4\pi) \partial_\mu \omega (x)$ 
leads to  a formal idealisation of a flux $\tilde G_{\mu\nu} (x)$ in the form of a tube in the equilibrium against the pressure of surrounding clouds of the SDM fields which are displace [18].

Now, we move on to the physics which operates with "large" and "small" systems. 
%Before to consider these systems, we 
Note that the success of cosmological nucleosynthesis calculations definitely gives the constraints to the existence of additional degrees of freedom that are in the thermo-equlibrium with the SM, or have an equal or greater temperature than the SM sector at the time period of nucleosynthesis.
The hidden (dark) sector will fall out of equilibrium with the SM once the Higgs sector has disappeared. 
The BS, as a "large" system, is supposed to be in equilibrium (the thermostat with $T$). The role of the "small" system is played by restricted region composed of the flux tubes (FT) defined by $N(R)$ number configurations of the FT of the length $R$. The stationary stochastic processes  are distorted by the random flux source $ \tilde G_{\mu\nu} (x)$ and by the weak action of a "large" system described by the SDM complex field $\phi (x)$ in the Lagrangian density (LD) $\sim {\vert (\partial_\mu - i g B_\mu )\phi\vert}^2$.
For simplicity, we only consider a minimal dark sector model containing a single Dirac fermion $\chi$, charged under a new gauge group $U^\prime (1)$, contributed to the formation of the BS. 
Like in the quark confinement in Nature, the energy would grow with the distance between $\bar\chi$ and $\chi$ because of  formation of an electric flux. The idea is that  $\bar\chi$ or $\chi$ may be a source (or a sink) of an electric flux which is expelled from the vacuum and is trapped in a thin FT connecting $\bar\chi$ and $\chi$. 
The excitation above the (classical) vacuum are the fluxes in the narrow tubes of a radius $\sim m_B^{-1}$ with $m_B$ being the mass of the field $B_\mu$. In the center of the FT the SDM condensate vanishes that is evident from the partition function for ensembles of systems with a single static FT
\begin{equation}
\label{e01}
Z_{flux} (\beta, R, m_B) = \sum_{\beta}\sum_{R} N(R) e^{-\beta E(R,\,m_B)} D(\vert \vec x\vert, \beta, M),
\end{equation}
where $N(R)$ is the main dynamical feature of the energy $E (R, m_B) \sim m_B^2\,R [a + b\ln (\mu_{IR} R)]$, $\beta = T^{-1}$. 
The $a$ and $b$  are real and positive constants, $\mu_{IR}$ is an infrared parameter. The function $D$ in (\ref{e01}) is associated with two-point correlation function of gauge-invariant operators at large distances [18]. The variable $\vert \vec x\vert$ in (\ref{e01}) can be replaced by the characteristic scale $L$ of the thermostat, with $L\sim $ the size of the BS $R_\star$. 
The mass $m_B$ develops an infinite fluctuation length $\xi\sim m_B^{-1}$ in the beginning of the BS formation at T close to the temperature of the CP, $T_{c}$. Actually, $ D = 1$ as $ T < T_{c}$. We consider $N(R)$ in the discrete space of a SDM condensate inside the BS
\begin{equation}
\label{e02}
N(R) = V\,l^{-3}\exp {\left [s \left (R/l\right)\right ]},
\end{equation}
where the FT with the entropy density $s$ lies along the links of a 3-dimensional cubic lattice of a volume $V$ with the lattice size $l << R$. Then, the partition function (\ref{e01})  has the form
%\begin{equation}
%\label{e03}
$$ Z_{flux} (\beta, R, m_B) = \frac{V}{l^3}\,\sum_{R} e^{-\beta E_{flux} (\beta, R)}, $$
%\end{equation}
where $ l\sim m_{\phi}^{-1}$ with $m_\phi$ being the mass of the SDM particle. Note that $m_B^2 (\beta)\sim g^2 (\beta) \delta^{(2)} (0)$, where $\delta^{(2)} (0)$ is the inverse cross-section of the FT, $\delta^{(2)} (0)\sim 1/(\pi r^2_s)$, $r_s \sim\xi$.
%= \left (\sqrt {2\lambda_\phi} \phi_0\right )^{-1}$, where $\lambda_\phi$ is the self-coupling constant, $\phi_0 = \langle \phi\rangle$ is the SDM order parameter.
The energy of the FT is $E_{flux} (\beta, R) = \sigma_{eff} (\beta) R$, where 
\begin{equation}
\label{e04}
\sigma_{eff} (\beta) = \sigma_0 - \frac{s}{l\beta} + \frac{L}{\beta R} M_{sc}(\beta).
\end{equation}
The sum of first two terms in (\ref{e04}) is the order parameter of the CP when the scale invariance is restored at $T_{c} = l\sigma_0/s $, $s = E_{tot}/T$. The total energy $E_{tot}$ is identified with the mass of the FT.   The last term in (\ref{e04}) describes the FT energy fluctuation at $T$ close to $T_{c}$ and higher with $M_{sc} (\beta)$ being the screening mass defined from the large distance exponential fall-off correlation function of the gauge operator at high $T$. The relation (\ref{e02}) counts the flux tubes that do not intersect the BS volume boundary. 
%The $\sigma_{eff}$ disappears at the temperature $T_{c} = l\sigma_0/s$. 
The FTs can be produced abundantly, and thus contributed much to the formation of the BS when the phase transition emerges at $T = T_{c}$.
The interaction range of the fields $\phi$ and $B_\mu$ forms the balance of the interaction between the flux tubes. It may be parametrised by the coefficient $k_{s} = \xi/l$ for the scale hierarchy and the expansion, where $\xi$ is approximately the radius of the FT, while $l$ stands for the coherent length of the SDM field. 
The time to the formation of the FT is $\tau_{FT}\sim O(1)\cdot\xi$ which leads to infinity at the CP.
While the attractive forces can appear between two parallel flux tubes at $k_{s} < 1$ (the FT radius $r_s\rightarrow 0$), the tubes repel each other when $k_{s} > 1$. The latter is suitable for the formation of the BS in the sense of evolving the star in size and increase with the mass. 

Since $k_{s}\lessgtr 1$, the oscillating character of the $B_\mu$ is assumed. The latter can be one of the constituents in a wide class of string-inspired models with dark forces in flux compactifications  [19]. If the hidden field $B_\mu$ exists, its oscillating contribution to BS formation with the direction around, e.g., the solar system is $\vec B = \tilde B\,\vec e_B \cos (x^0/\xi)$, 
where $\tilde B$ is the amplitude of the oscillation, $\vec e_B$ is the unit vector pointing to the direction of $\vec B$. Note, that for very light $B_\mu$ field, there will be no oscillations, and the direction with $\vec e_B$ is the only main input to search for BS expansion with increasing of the amplitude $\tilde B$. The effective electric field induced by the $B_\mu$ oscillation is 
$\vec E =  \sqrt {2\rho_B}\, \vec e_B \sin (x^0/\xi)$, where $\rho_B = \rho_\odot - \rho_\star $. Here, the amplitude $\tilde B$ is related with the density $\rho_B$ of the hidden $B_\mu$ boson by $\rho_B = (1/2) m_B^2\,\tilde B^2$.
%$\rho_\odot$ is the DM density in the vicinity of the solar system. 
The $\rho_\star$ approximated by the quantity $\rho_\star = M_\star n_\star + m_\phi n_\phi (1 + \delta_h)$ gives the density of the BS with the density number  $n_\star$, the density of the SDM (the second term) with the contribution from the Higgs bosons inside the star defined by $\delta_h \sim\left (\tau_h/\tau_\phi\right ) < 10^{-14}$, where $\tau_h$ and $\tau_\phi$ are the lifetimes of the Higgs boson and the SDM particles, respectively [11]. 
Here, the lower bound on $\tau_\phi > 1.3\cdot 10^{-7}$ sec  from the collider's stringent limits of the branching fraction of the Higgs boson decay to two long-lived particles  subsequently decaying to $\tau^+\tau^-$ pairs [6], has been used.   On the other hand, the constraints on $\tau_\phi$ in study the cosmic microwave background may be close to the age of the Universe ($\tau_{univ}\sim 10^{13}$ sec at recombination) or may exceed  $\tau_{univ}$ in case of the galactic diffuse emission for MeV-GeV DM annihilation primarily to $e^+e^- $ pairs, or to particles, e.g., the SDM, which decay dominantly to $e^+e^-$  final states [20]. The latter may almost annul the Higgs bosons contribution completely to the formation of the BS,  $\delta_h < 10^{-34}$.

%he decay of $\phi$ to SM particles by requiring  the $10\%$ DM subcomponent ($\Omega_\phi = 0.1 \Omega_{DM}$) in study the cosmic microwave background leads to the more %strong constraint for the scalar LLPs, $\tau_\phi \geq  10^{23}$ sec [17]. The latter may almost annul the Higgs bosons contribution completely in the BS formation,  $\delta_h < %10^{-44}$.
 
For large enough $\xi >> x^0$ $(k_{s} > 1)$, the electric field disappears, the FTs repel each other, and the BS undergoes the un-fluctuated expansion (with no oscillations in time). 
Note that all the discussion above concerning the formation of the BS in terms of the FTs is valid for the characteristic time $\tau_B\sim\xi > x^0$
%The $\vec E$ field arises only if $x^0 < \xi$, 
where $\xi$ is estimated to be $\xi \sim0.4\cdot 10^{-3} \tau_h/v_B^2$. The Higgs boson contribution to the BS formation is efficient if the hidden $B_\mu$ boson with the mass $m_B < m_\phi\sim m_h/2$ has the velocity $v_B \leq 10^{-3}$ in the medium (inside the BS).
The BS is unstable, and the SDM inside may decay dominantly into two dark photons (DP), each subsequently decaying into a pair of the SM particles via kinetic mixing. The final states may contain the charged tracks, which can be electrons, muons or light hadrons, and also the neutrinos as a missing energy.

\section{(Thermo)dynamics with the Higgs portal. The BS mass.}

A cosmological level separating the high from low energy density DM fields may be scaled with the energy density of the  BS, $\rho_\star << 10^{-10} \rho_\odot $ [11]. For low $\rho$ a reasonable description of a system composed by SDM + SM fields would be obtained by considering it as a gas of free scalar particles. There is no experimental evidence for existence of high energy density macroscopic cosmological objects, like the BSs, of nearly non-interacting scalar fields inside the star. The (thermo)dynamic relation for the total energy $E_\star$ of the BS is 
%\begin{equation}
%\label{e15}
$ E_\star = -PV + s/\beta + \mu N_\mu, $
%\end{equation}
where $P$ and $V$ are the pressure and the volume, $s$ is the entropy, $N_\mu$ are the conserved quantum numbers of the system with the associated  chemical potentials $\mu$. The pressure is positive if both $T$ and $\vert\mu\vert$ are large enough. If the theory binds the scalar fields into the BS at large distances, then at low $T$ and $\vert\mu\vert$ one can assume $P < 0$. The latter may be interpreted as strongly binding SDM particles to the interior of the BS. As a consequence, there exists a set of critical values $\beta_c$ and $\mu_c$ such that $P(\beta_c, \mu_c) = 0$. The level of density of the BS rises exponentially as energy is fed into the star, $\sim exp\left [(E_\star - \mu  N_\mu)\beta\right]$. If $E_\star$ is held fixed, the BS is in equilibrium and hence may be in the (meta)stable state: an expansion leads to $P < 0$, while the contraction is for positive pressure. 
%The BS will be unstable with respect to photons, DPs emission because of the decays of the SDM particles, however. 
Thus, one can interpret the BS as a large scale and high mass resonant (thermo)dynamical bound state of the SDM in the Universe.

An effective interaction between the singlet spinor DM field $\chi (x)$, the real SDM field  $\phi (x)$ and the SM in terms of the Higgs doublet $ H^+ H(x)$ is given by the LD
 \begin{equation}
\label{e9}
 L\supset y_\chi\hat m\bar\chi\chi + \frac{1}{2}\partial_\mu\phi\,\partial^\mu\phi - \Omega, 
\end{equation}
where $y_\chi$ is the real Yukawa coupling constant, $\hat m = m\, \phi/\phi_0 = m\,\varphi$ with $m$ being the mass of $\chi$-particle. The $\phi_0 =\langle \phi\rangle$ is the only dimensional model order parameter of the scale symmetry breaking which arises after the phase transition. The DM $\chi$ enjoys a global $U(1)$ symmetry that ensures its stability. Below the phase transition the Yukawa interaction leads to the mass $m\sim y_\chi \phi_0 > T_{c}\sim m_\phi$. 
The (thermo)dynamical potential (ThDP) $\Omega$  in (\ref{e9}) is
\begin{equation}
\label{e10}
 \Omega = V(\phi, {\vert H\vert}^2) + \Omega_D + \Omega_{T=0},
\end{equation}
where the particular form of the potential $V(\phi, {\vert H\vert}^2)$ gives $\phi$ a mass $m_\phi$; $\Omega_D$ is the dark matter ThDP.
%, while $\Omega_{T=0}$ is added so that $\Omega = 0$ at zero temperature where the Higgs portal smears. 
The Higgs portal through the interaction (\ref{e10}) with the Higgs doublet ${\vert H\vert}^2$ ensures the communication of the dark sector with the SM [21,22]. The $\Omega_{T=0}$ in (\ref{e10}) is added so that $\Omega$ in (\ref{e9}) is zero at $T = 0$ where the Higgs portal smears. 
The Higgs portal allows the SDM particles to decay to SM particles that can be realised in the BS because of its instability. The potential $V$ in (\ref{e10}) has to be invariant under the $\mathbb {Z}_2$ background symmetry and may posses an explicit scale symmetry breaking where the source of this breaking is an additional operator, e.g., related to SDM field with scaling dimension $d\neq 4$ [23]. In the approximate conformal field theory (CFT) when $\vert d-4\vert <<1$, the entire potential for $\phi$, up to correction of order $O\left [(d-4)^2\right ]$, is
\begin{equation}
\label{e11}
 V(\phi, {\vert H\vert}^2) = \frac{1}{4}\kappa_D\phi^2\,{\vert H\vert}^2 + \frac{\lambda}{4}\,\phi^4 \left (\ln\frac{\phi}{\phi_0} - \frac{1}{4}\right ),
\end{equation}
where $\kappa_D \leq O(1)$, $H = (v + h)/\sqrt 2$, $\langle {\vert H\vert}^2\rangle = v^2/2$ with $v$ being the vacuum expectation value (VEV) of the Higgs doublet ${\vert H\vert}^2$ after EW symmetry breaking; $h$ is the SM Higgs boson field. The self-coupling constant  is 
%\begin{equation}
%\label{e12}
 $\lambda = {\left ({m_\phi}/{\phi_0}\right )}^2 (1- \delta_D), $
%\end{equation}
where  $m_\phi$ is given by $d^2 V/d\phi^2$ minimised at $\phi_0$ and $\langle {\vert H\vert}^2\rangle$, $\delta_D = \kappa_D (v/2m_\phi)^2$. Conformal invariance of the DM scalar sector can be broken at a scale $\Lambda_{CFT}\sim 4\pi\phi_0 >   4\pi v$. A non-zero expectation value for $\phi$ implies the presence of a SDM condensate. The constant $\lambda$ depends on $\phi_0$, $\kappa_D$ and $m_\phi$ where the latter is a free parameter that can be viable in a wide mass range from MeV to TeV. Based on the CMS data  [24] on the DM interactions through the Higgs portal, the mass range for $m_\phi$ is
$ [0.1\, GeV,\,m_h/2]$, where $m_h$ is the SM Higgs boson mass.
Taking into account the relation for DM relic density $\Omega_D h^2 = 0.1199\pm 0.0027$ [25] required for relic density of the SDM $\phi$-particles which are totally produced by the thermal freeze-out process, the value for $\kappa_D$ is minimal, $\sim O(10^{-4})$, at $m_\phi = m_h/2$, where the annihilation cross section is resonantly enhanced [26]. On the other hand, $\kappa_D$ tends to $\sim O(1)$ for $m_\phi > 1$ TeV which is not favoured neither by the EW vacuum stability, nor the analysis with the LHC data.  If we use the CMS data [24] for the constraint of the branching ratio of the Higgs boson invisible decay $Br (h\rightarrow inv) <$ 0.15 at 95 $\%$ C.L., one can estimate the upper limit $\kappa_D < $ 0.034 for $m_\phi = $ 62 GeV. Here the invisible decay width is given by 
%\begin{equation}
%\label{e13}
 $$\Gamma (h\rightarrow \phi\phi) = \frac{\kappa_D^2\,v^2}{32\pi m_h}\sqrt {1- {\left (\frac{2 m_\phi}{m_h}\right)}^2}, $$
%\end{equation}
 where the Higgs boson decay width with the SM expectation of $\Gamma^{SM}_h = 4.1 $ MeV and the $\Gamma_h = 3.2 + 2.4 - 1.7$ MeV measured by CMS [27] are used. 
The upper limit estimated for $\kappa_D$ is in agreement with the Higgs portal coupling constraint  $\kappa_D \leq 0.028 { [1 - { ({2 m_\phi}/{m_h})}^2 ]}^{-1/4} $ if $m_\phi < m_h/2 \simeq 62.5 $ GeV [28], whereas $\kappa_D$ is almost unconstrained if $m_\phi > m_h/2$.
At the EW scale with $\phi_0\sim O(10^3 GeV)$ [29]  one has $\lambda\sim O(10^{-4})$, that would imply the 
vacuum energy density 
\begin{equation}
\label{e14}
 \vert\rho_v\vert \simeq \frac{m_\phi ^4}{16\lambda} (1- 6\delta_D)\sim O(10^{50} GeV\,cm^{-3}).
\end{equation}
The result (\ref{e14}) is almost the same as those of the upper limit on the  characteristic energy density inside a $\lambda = 0$ configuration, $\rho \sim (1/8)\kappa_D \phi^2_0\,v^2 < 2.5\cdot 10^{50} GeV cm^{-3}$.

The ThDP of an open cosmological system containing the $\chi$ DM and the SDM fields is
\begin{equation}
\label{e166}
\Omega_D = -\frac{1}{\beta\,V}\ln\left (Z_0\cdot Z [\varphi, \mu]\right ),
\end{equation}
where $\Omega_0 = {\left (\beta\,V\right )}^{-1} \ln Z_0$ is the ThDP with partition function $Z_0$ in the absence of interactions (the coupling constants set to zero). The interaction part of (\ref{e166}) with the partition function $Z$ is 
%\begin{equation}
%\label{e17}
$$Z[\varphi,\mu] = \int D\bar\chi D\chi \,exp \left [ -\int_0^\beta d\tau\int_V d^3 \vec x\,\bar\chi\left (p_\mu\gamma^\mu - \hat m\right)\chi + \mu\bar\chi\chi\right ]. $$
%\end{equation}

The SDM sector and the SM are thermally decoupled. The temperature of the dark sector is different than that of the SM and it also cooler than the SM since the DM states freeze-out in the SM sector. It allows one to assume that the reduced SDM field $\varphi$ is an equilibrium field $\tilde\varphi$ in medium. Carrying out the minimisation of the ThDP (\ref{e10}) one can find the Eq.
%\begin{equation}
%\label{e18}
$$\tilde\varphi^2\,\ln\tilde\varphi = \frac{1}{\lambda\,\phi_0^2} \left \{{\left (\frac{2 m}{\phi_0}\right )}^2 \left [F(\beta,\mu) + (\mu\rightarrow -\mu)\right ] - \frac{1}{4}\kappa_D\,v^2\right \}, $$
%\end{equation}
where 
%\begin{equation}
%\label{e19}
$$ F(\beta,\mu) = \int \frac{d^3 \vec p}{(2\pi)^3} \frac{n_p}{\tilde E_p} $$
%\end{equation}
with $n_p = {\left [1 + exp(E\,\beta)\right ]}^{-1}$, $E = \tilde E_p - \mu$, $\tilde E_p = \sqrt {p^2 + \tilde m^2}$,  and $\tilde m = m\tilde\varphi$.
Note that the $\chi$ DM contribution is made out of separate $\chi$ and $\bar\chi$ DM fields because of the finite chemical potential. 
If the DM sector is thermalised with the SM sector being in an almost plasma state in the early Universe, the cosmological expansion which causes this plasma to cool  will make the interactions between the SM with the DM rather weak, driving both these sectors out of equilibrium.
Whether the SDM actually dominates the Universe depends on its fluctuations.
%We will see that the fluctuations of $\tilde\varphi$ with $T$ are visible at temperatures $\sim O(v)$, at the weak scale. 
For low values of the chemical potential $\mu << \tilde E_p$, the fluctuations of the field $\tilde\varphi$ with $T$ around its equilibrium state are visible at the weak scale and are suppressed at $T\rightarrow 0$:
\begin{equation}
\label{e20}
\tilde\varphi (\beta)\simeq 1 + \frac{1}{\lambda\,\phi_0^2} \left [{\left (\frac{2 m}{\pi}\right )}^{{5}/{2}}\,\frac{2}{\phi_0^2\,\beta^{{3}/{2}}} e^{- m\,\beta} - \frac{1}{4}\kappa_D v^2\right ].
\end{equation}
%The field $\tilde\varphi$ can potentially form the Bose-Einstein condensate (BEC) if the temperature is sufficiently low, $\tilde\varphi (\beta\rightarrow \infty)\rightarrow 1$ as $\kappa_D\rightarrow 0$ (no interaction with the Higgs). 
The DM can exist at $T\simeq 0$, thus leading to its total condensate ($\kappa_D =0$).
%under the condition for the SDM mass in medium
%\begin{equation}
%\label{e21}
%m_\phi^2 > {\left (\frac{2 m}{\pi}\right )}^{\frac{5}{2}}\,\frac{2}{\phi_0^2} \,T^ {\frac{3}{2}}\,e^{- m\,\beta} .
%\end{equation}

Now we move on to the question of the BS mass and the role of the interactions between the SDM and the SM inside the star. 
The spinor DM $\chi$ is in thermal equilibrium with both the SDM sector and the SM sector at the temperature of the phase transition $T_{c}$. The $\phi$-fields undergo the phase transition of the first order, where $\langle\phi\rangle$ instantly approaches the value $\phi_0$ and varies little with $T$ around $T_c$. 
%We treat $m$ and $m_\phi$ as the constants.
The phase transition proceed through the binding of the SDM and the SM sector fields and growth of the scalar mixed states or the flux tubes at $\langle\phi\rangle = \phi_0$. The latter states expand and merge into a BS under the gravitational forces and the interactions between the fields (including the self-interactions), until the Universe has evolved and transitioned.  One of the critical roles in the formation of the BS as the bound state of the SDM fields through the Higgs portal belongs to interaction energy between the fields. Based on the Heisenberg principle of uncertainties, if there are no interactions between the fields inside the star ($\lambda\rightarrow 0, \kappa_D\rightarrow 0)$, the BS is in equilibrium as a relativistic bound state with the idealised mass $M_\star^{id} = M_{Pl}^2/m_\phi$. The effect of interaction and the binding between the fields can be estimated on the potential level (\ref{e11}) compared to the SDM energy density of non-interacting  fields, $\rho = m_\phi^2\,\phi^2$. For small enough $\lambda$ and $\kappa_D$ the relativistic energy density is proportional to
\begin{equation}
\label{e202}
\sim \frac{1}{2 (2m_\phi)^2} \left (\kappa_D\,v^2 + \xi_\phi\,\lambda_{eff}\,M_{Pl}^2\right ),
\end{equation}
where $\phi^2\sim \xi_\phi\,M_{Pl}^2$ inside the relativistic BS, $\lambda_{eff} = \lambda\, [\ln (\sqrt {\xi_\phi}\,M_{Pl}/\phi_0) - 1/4]$. The constant $\xi_\phi\sim O(10^{-30})$ is defined from the physically relevant condition $\Delta_{eq} \leq 1$ when the families of an equilibria between the fields interactions and the gravity is parametrised by $R_{eq} = 1 + \Delta_{eq}$ with $\Delta_{eq} $ being 
$$\Delta_{eq} \simeq \frac{\xi_\phi\,\lambda_{eff}\,M_{Pl}^2}{2 (2m_\phi)^2}\left ( 1 + \frac{1}{2}\delta_D\right ). $$
% \left [ 1 + 2\kappa_D {\left (\frac{v}{4 m_\phi}\right )}^2\right ].$$
The upper limit of the BS mass $M_\star^{max}$ differs from that of the idealised one  $M_\star^{id}$ with the temperature and the spinor DM mass $m$
%, the coupling $\kappa_D$ and the self-coupling $\lambda_{eff}$ 
\begin{equation}
\label{e22}
%M_{\star}^{max} < \frac{1}{\sqrt {2}}{\left ( \frac {\pi}{2 m}\right )}^{5/4}\, \frac{M_{Pl}^2\, \phi_0}{T^{3/4}}\, K_{eq}\, e^{m/2T}.
M_{\star}^{max} \simeq \frac{1}{\sqrt {2}}{\left ( \frac {\pi}{2 m}\right )}^{5/4}\, \frac{M_{Pl}^2\, \phi_0}{T^{3/4}}\,R_{eq}\, e^{m/2T}.
\end{equation}
%Here,  the equilibrium between the field interaction forces and the gravity inside the BS is governed by unified dimensionless quantity (see (\ref{e11})) 
%$$ K_{eq} \simeq \lambda_{eff}\,{\left (\frac{M_{Pl}}{2 m_\phi}\right )}^2 \,\left [1 + 2\kappa_D {\left (\frac{v}{4 m_\phi}\right )}^2\right ], $$
%$\lambda_{eff} = \lambda\, [\ln (M_{Pl}/\phi_0) - 1/4]$, $m_\phi > \kappa_D\,v/(2\sqrt 2)$. 
Taking into account the constraint on the mass of thermally produced DM $\leq 100$ TeV [30-32],  
%for $m \sim 10$ TeV, $T\sim T_{c}\sim m_\phi$, 
one can find $ M_{\star}^{max} \sim O(10^{70} GeV)$ that is astro-physically relevant, e.g., for $m \sim 10$ TeV, $T\sim T_{c}\sim m_\phi$. The maximal mass (\ref{e22}) is the thermal dynamical indicator for the stable and unstable configurations  of the star. 
%The latter is characterised by $T_c$ where 
The energy density of the DM spinor $\chi$-particles inside the star with equilibrium abundance is bounded by 
%\begin{equation}
%\label{e221}
$\rho_\chi = m n_\chi < \pi {\left ({ m_\phi\,\phi_0}/{4}\right )}^2. $
%\end{equation}
Here, the number density of $\chi$-particles entered the star is 
%\label{e222}
$n_\chi = s_\chi  {\left ({ m\, T_c}/{2\,\pi}\right )}^{3/2}\,e^{- m/T_c}$
%\end{equation}
with $s_\chi = 2$ being the spin states number. The result for $\chi$ DM energy density  fed into BS, $\rho_\chi \leq 10^{51} GeV\,cm^{-3}$, is comparable with those of (\ref{e14}) for the vacuum energy density at finite $\lambda\sim O(10^{-4})$.
The BS nucleates wide at $T_c$ where the size is proportional to 
\begin{equation}
\label{e223}
R_\star (T_c) \sim \frac{m^{5/4}}{M_{Pl}^2\,\phi_0}\, T_c^{3/4}\,e^{- m/2\,T_c},
\end{equation}
%with $ T_c < m/2$, 
then shrinks and becomes dense as it accumulates the scalar DM particles interacting with Higgs bosons as $T < T_c$. The lower bound on $T_c$ is given by 
%\begin{equation}
%\label{e224}
$$T_c > {\left (\frac{M_{Pl}^2}{M_\star^{max}}\right )}^2\,\frac{1}{m_\phi}\,N^{2/3}\sim 10^{-66}\,N^{2/3} GeV $$
%\end{equation}
for $N$ gravitationally interacting light SDM $\phi$ - bosons in connection with large occupation numbers $n_f$ in thermal equilibrium
%\begin{equation}
%\label{e225}
$$\sum_f n_f = \sum_f \frac{1}{\bar\rho_0^{-1} e^{F(f)\beta} -1} = N  $$
%\end{equation}
to formation of a Bose-Einstein condensate. Here, the relative (dimensionless) energy density $\bar\rho = \rho_\star/\rho$ is the stationary stable value at $\bar\rho =\bar\rho_0$ when the BS is formed against its energy density $\rho_\star$ with $F(f) = E(f) - \mu\,Q(f)$, where $E(f)$ is the energy, while $Q(f)$ stands for the conserved charge. 
%The number of particles $ N < O(10^{105})$ for $m \leq O(100 \,TeV)$ and  $ N \sim O(10^{102})$ for $m \sim O(1\, TeV)$.
The large number $N$ is expected when $\bar\rho_0\rightarrow 1$ at high $T$ and when the SDM particles are light. 
If  $ T_c < m/2$, the number of particles $ N < O(10^{105})$ for $m \leq O(100 \,TeV)$ and  $ N \sim O(10^{102})$ for $m \sim O(1\, TeV)$.
For the heavier SDM bosons ($m_\phi > T_c)$, the condensates are smaller with the characteristic size and the mass of the BS as an astrophysical exotic compact object of DM that is neither the neutron star, nor the black hole.

\section{The BS formation. The model with gravity.}
The probability to formation of the BS depends on the amplitude of perturbations, where in  case of the complex SDM fields these initial perturbations can cause the gravitational instability. The latter may be associated with deviations from the spatial uniformity related to the presence of the flux tubes.
We assume the field  sector in the BS is composed of the dark sector with the complex scalar field $\phi(x)$ that communicates with the hidden vector field $B_\mu (x)$ in the LD $L(x)$ that combines also the gravity (GR) and the SDM sector itself:
\begin{equation}
\label{e141}
\frac{L}{\sqrt{-g}} = \frac{\mathcal{R}}{16\pi G}  - \frac{1}{2} g^{\mu\nu}\partial_\mu\phi\,\partial_\nu\phi^\star + L_D.  
\end{equation}
Here, the first term is the Einstein-Hilbert action for GR with $\mathcal{R}$ being the Ricci scalar for background metric $g_{\mu\nu}$;  $G = M_{Pl}^{-2}$. 
%The DM Abelian Higgs LD has the form
The $L_D$ in (\ref{e141}) contains the fields $\phi(x)$ and  $B_\mu (x)$ with the generating current $k_\mu = \partial^\nu B_{\mu\nu}$.
% $B_{\mu\nu} = \partial_\mu B_\nu -  \partial_\nu B_\mu$. 
Before we define $L_D$, it is necessary to make some digression. Let us consider an operator $Q_R$ related to $k_\mu$ relevant to BS in $S(\Re^4)$ space, 
 \begin{equation}
\label{e1420}
Q_R = \int k_0 (\vec x)\, t_R (\vec x)\, d^3 \vec x,
\end{equation}
where the test function $t_R (\vec x)$ = const for all the points $\vert \vec x\vert \leq R_\star$ inside the star of the size $R_\star$. In some sense, the (\ref{e1420}) is the "charge" that generates locally the group of automorphism  $U(1)$ corresponding to gauge transformations to the first kind [33]. From the physical point of view, for any field $X(x)$ (it may also be $\phi (x)$) with respect to $k_\mu (x)$ we have the "charge" $q$, if
\begin{equation}
\label{e1421}
\left [Q_R, X\right ] = \left [\int \partial^i B_{0i} (\vec x)\, t_R (\vec x)\, d^3 \vec x, X(y) \right ] = - q\, X(y)
\end{equation}
for $t_R \in S(\Re^4)$ at any (large enough) $R_\star$. On the other hand, for local $k_\mu$ where $B_{\mu\nu} = - B_{\nu\mu}$, for any field $X(x)$ local relative to $k_\mu$ and $B_{\mu\nu}$, one can conclude the (\ref{e1421}) is zero. In order $k_\mu$ generates the non-trivial automorphism on the local algebra, we have introduced the "phantom" local field $P_\mu (x) = (\theta -1) \partial_\mu\partial_\nu B^\nu (x)$ in the sense that the standard term $\sim B_{\mu\nu} B^{\mu\nu}$ in  $L_D$ in  (\ref{e141}) is distorted by the $\theta$-term, $(\theta/2) \partial_\mu B^\nu \partial ^\nu B_\mu$ with the real number $\theta \neq 1$. The $L_D$ is
\begin{equation}
\label{e142}
L_D = -\frac{1}{2} {\left (\partial_\mu B_\nu\right )}^2 + \frac{\theta}{2}\partial_\mu B_\nu\partial^\nu B^\mu + {\left \vert \left (\partial_\mu + i g B_\mu \right )\phi\right \vert}^2 - \alpha {\left ( {\vert \phi\vert}^2 -\phi_0^2\right)}^2 - f {\vert \phi\vert}^4,
\end{equation}
where  $g$ is the coupling constant; $\alpha > 0$, $f  > 0$.  
%The kinetic term in (\ref{e142}) contains a real number $\theta\neq 1$ in order to avoid the trivial relation for the generating current $k_\mu = \partial^\nu B_{\nu\mu}$  with any localised %random field $X$, $\left [\int d^3\vec x\, k_0(x), X\right ] = 0$, in case the locality in the BS is valid and one can not measure both $k_0 (x)$ and $X$ simultaneously (see also [27] for %local QFT where the unphysical local field $\mathcal{A}^\mu (x) =\partial_\nu F^{\nu\mu}(x) - k^\mu (x)$ is introduced).
%The cosmological constant is associated with 
The last term in (\ref{e142}) is associated with the cosmological constant, where $f$ corresponds to the existence of a flat direction regarding the potential in $L_D$ if $ f = 0$.
The LD (\ref{e142}) is invariant under transformations
%\begin{equation}
%\label{e143}
$B_\mu (x)\rightarrow B_\mu(x) +\partial_\mu\Lambda (x), \phi(x)\rightarrow e^{-i g\Lambda (x)}\,\phi (x), $
%\end{equation}
where $\Lambda(x)$ is an any real function obeying the Eq.  $\Box \Lambda (x) = 0$. We suppose that an evolving Universe at its early stages was driven by the SDM fields that were minimally coupled to GR in the sense of dynamical fields. In the BS, the main contribution comes from the strong DM scalar sector where $M_{Pl}^2 << {\vert\phi\vert }^2$ may be the dynamical quantity. This corresponds to the following LD
\begin{equation}
\label{e144}
\frac{L}{\sqrt{-g}} = \frac{1}{2} \mathcal{R}\,\zeta_\phi\, {\vert\phi\vert }^2  - \frac{1}{2} g^{\mu\nu} \partial_\mu\phi\,\partial_\nu\phi^\star + L_D, 
\end{equation}
where $\zeta_\phi$ is a new parameter, which defines the cutoff of the theory, $\Lambda_{cut} \sim M_{Pl}/\sqrt {\zeta_\phi}$. The important properties of (\ref{e144}) are the scale invariance and the dynamical behaviour of the Planck scale, $M_{Pl}^2 \sim \zeta_\phi\, {\vert\phi\vert }^2$. 
We find the field equations
 \begin{equation}
\label{e145}
\left [{\left (\partial_\mu + i g B_\mu\right )}^2 - \frac{1}{2}\partial_\mu\partial^\mu\right ] \phi = 2 (\alpha + f) \left (\phi_0^2 - {\vert\phi\vert}^2 \right ) \phi,
%{\left (\partial_\mu + i g B_\mu\right )}^2\phi = 2 (\alpha + f) \left [ \phi_0^2 - {\vert\phi\vert}^2 + \frac{1}{4 (\alpha + f)} \Box\right ] \phi,
\end{equation}
 \begin{equation}
\label{e146}
\tilde k_\mu = k_\mu + P_\mu \equiv - \Box B_\mu  + \theta \partial_\mu (\partial\cdot B) = 2 g^2 B_\mu {\vert\phi\vert}^2 - i g \left (\phi^\star\partial_\mu \phi - \phi\,\partial_{\mu}\phi^\star\right ).
\end{equation}
Note, that the parameter $\zeta_\phi$ does not enter the Eq. (\ref{e145}) because of the 
%To find the Eq. (\ref{e145}) the inclusion of GR, leading to appearance with the 
additional ground state $ \mathcal{R} = 4 f\phi_0^2/\zeta_\phi$ at  $f\neq 0$.
The solution of the Eq. (\ref{e146}) is 
\begin{equation}
\label{e147}
g  B_\mu  = \frac{1}{2 g}\frac{\bar k_\mu}{\varphi^2} - \partial_\mu\eta,
%g \left (B_\mu + B_\mu^{sing}\right ) = \frac{1}{2 g}\frac{k_\mu}{\varphi^2} - \partial_\mu\eta,
\end{equation}
where the polar decomposition of the SDM field is used, $\phi (x) = \varphi (x) e^{i\eta (x)}$.
The electric flux can be quantised if the proper boundary conditions of Eq.  (\ref{e146}) are known. The flux is 
\begin{equation}
\label{e149}
\Phi = \int B_{\mu\nu} \,d\sigma^{\mu\nu} = \oint  B_{\mu}(x)\,dx^\mu,
\end{equation}
where $\sigma^{\mu\nu}$ is a two-dimensional surface element $\sim \Delta x^\mu\Delta x^\nu$ $(\mu\neq\nu)$ in $S(\Re^4)$. 
The flux (\ref{e149}) guarantees the nonobservability of the FT when the  gauge symmetry is not broken.
At the cosmological scale, the contribution from $k_\mu$ is negligible if we integrate out over a large closed loop in (\ref{e149}). The rest loop integral contains two terms
\begin{equation}
\label{e150}
\Phi \simeq \frac{\theta -1}{2 g^2} \oint \partial_\mu \left (\partial\cdot B \right)\frac{1}{\varphi^2}\,dx^\mu - \frac{2\pi n}{g},
\end{equation}
where the "phantom" field contribution in the first term would be neglected because $\phi^2\sim O(10^{-30}) M_{Pl}^2$ inside the relativistic BS (\ref{e202}), or even  $\vert \phi\vert^2 >> M_{Pl}^2$ being assumed for the dynamical character of the quantity $\vert \phi\vert^2$ to formulate the LD (\ref{e144}). The second term in (\ref{e150}), under the assumption that the phase $\eta (x)$ in (\ref{e147}) is that the field $\phi(x)$ should be a single valued, is the result of variation with the winding integer number $n$ (the "topological charge" of the FT) by $2\pi n$ over a large closed loop. Thus, the flux (\ref{e150}) is quantised as $\Phi\simeq -2\pi n/g$, resulting from the formal feature, that $\Phi$ should not be defined more precisely than that of $2\pi$ times $n$.

The BS is composed of a large number of FTs which are distributed inside the star under the requirement of translational invariance within the SDM condensate configuration. The latter is governed by the field equations  (\ref{e145}) and (\ref{e146}) with an appropriate coordinate symmetry. For the FT as the field topological object extended in the space, the cylindrical symmetry with the coordinate $r_s$  (the radial distance from the center of the FT) is most suitable, where $\varphi = \varphi (r_s)$ and $B_\mu\rightarrow \vec B = [\bar B (r_s) / r_s]\vec e_{\theta_{az}}$ with $\vec e_{\theta_{az}}$ being the unit vector. The field phase $\eta = n \theta_{az}$, where $\theta_{az}$ is the azimuth angle around $z$ axis, while $\vec\nabla \eta = (n/r_s)\vec e_{\theta_{az}}$. The rotation of $B_\mu$ field gives the electric field $\vec E = E_z(r_s) \vec e_z$ with the unit vector $\vec e_z$ along the $z$ axis.
%In the early stage of the BS formation, $\varphi (r)\sim 0$, $\bar B (r)\sim 0$ as the FT radius $r\rightarrow 0$. 
The essential point for the BS formation in the early Universe is that there will be the singularity at the center of the flux which comes from the quantisation  condition $\vec \nabla\times \vec\nabla \eta= (2\pi n) \delta (x)\delta (y) \vec e_z$. 
Since the phase $\eta$ is an arbitrary function with $n$, the latter is free to grow up as winding number. When it is happened, the flux has to grow as well as the characteristic size of the FT.
There are the repel forces between the parallel flux tubes where $k_s = (\sqrt{2\alpha}/g) > 1$ for the masses $m_\phi = 2\sqrt{\alpha}\phi_0$ and $m_B = \sqrt{2} g\phi_0$ of the SDM and the $B_\mu$ fields, respectively. That is true because of increasing the characteristic distance of the interaction $\sim 1/g$. The quantity $\vert \Phi\vert$ has to grow as the self-coupling constant $\alpha\rightarrow 0$ and $n\rightarrow\infty$, $\vert \Phi\vert > \sqrt{2/\alpha}\, \pi n$.
If one wants to have the BS being visible in terms of huge numbers of the FTs, all is need to do is to look for solution with the singular structure of the SDM field through its phase.
The evolving dynamics of the BS is when the coherent length $l$ of the large $\phi$ is much smaller than the characteristic size $\sim \xi$ of the FT up to the vicinity of the value when the singularity in the phase  $\eta (x)$ can arise at the core of the FT.  
The analytical solution for the $B_\mu$ field configuration inside the star during its evolving in $r_s$ direction can be obtained in the SDM condensate approximation $\varphi (r_s)\rightarrow \phi_0$ at large distances $r_s >>  l\sim  m_\phi^{-1}$. The use of this approximation leads to the following equation from   (\ref {e146})
%\begin{equation}
%\label{e16}
$$ \left (\frac{d}{d r_s} - \frac{\theta}{r_s}\right ) \frac{d\bar B(r_s)}{dr_s} = 2 g^2 \phi_0^2 \left [\frac{1}{2\pi} \Phi + \bar B(r_s)\right ]. $$
%\end{equation}
%where $\bar\phi_0^2\simeq \left [1 +\xi_\Box /(\alpha + f)\right ]\phi_0^2$ and $\xi_\Box$ is the characteristic coefficient after assuming that the scales related with the metric %$g^{\mu\nu}$ be dynamical in the LD and these scales would be replaced by the large SDM fields $\phi$ in the equation of motion, $1/2\partial_\mu\partial^\mu\rightarrow 2\,\xi_\Box %{\vert\phi\vert}^2$. 
The results to the profiles of the vector field $\bar B(r_s)$ and the electric field $E_z(r_s)$ would be as
% \begin{equation}
%\label{e17}
%\bar B (r_s) \simeq - \frac{1}{2\pi} \Phi - \sqrt{r_s m_B}\, e^{-\kappa}, \,\,\, E_z(r_s) \sim \frac{\sqrt {m_B}}{r_s^{3/2}} e^{-\kappa}
$$ \bar B (r_s) \sim - \frac{1}{2\pi} \Phi - \sqrt{\bar r}\, e^{-\bar r}, \,\,\, E_z(r_s) \sim \frac{m_B^2}{\bar r^{3/2}} e^{-\bar r} $$
%\end{equation}
for large values $r_s $ outside the region with the singularities at the core of the FT with $\bar r = m_B\,r_s$.
%$\kappa = 1-\theta +\sqrt { \theta^2 + 2\left (1+ 4g^2\,\bar\phi_0^2 \,r_s^2\right )}$.
%In case $\theta = 1$ (no "phantom" influence), $\varphi (r_s)\sim 0$, $\bar B (r_s)\sim 0$ as the FT radius $r_s\rightarrow 0$. 
%On the other hand, 
During the evolving of the BS, the SDM field increases as $\varphi (r_s)\rightarrow \phi_0$, the vector field with the "topological charge" $n$ approaches the ratio $\bar B(r_s)\rightarrow n/g$ as the BS increasing with the size, $r_s\rightarrow\infty$. The FT solution appears as the topological excitation as the relevant collective mode in the vacuum that provides intuitive picture of the stars in terms of the FT structure of the electric flux. The BS can also be regarded as the massive astronomical FT (MAFT) with no end points (the "terminals"), the co-called "the MAFT ring" excitation.  The latter is the subject for next studies.

\section{Conclusion}
In conclusion, the study of feebly interacting dark matter fields with the SM fields, which are contribute to the BS formation, is presented. 
In particular, the formation of the star may be explained in terms of the electric fluxes of $B_\mu$ field under the influence of the SDM field as the cosmological dynamical quantity minimally coupled to GR. 
%We establish that the scalar dark matter (SDM) massive fields can form the macroscopic scalar bound state when minimally coupled to gravity. 
The star may have approximately a stationary form with the maximal mass  larger that the solar mass. 
%We have also considered an effective description of a more complete model with the (thermo)dynamic potential of the BS at finite temperature for an arbitrary number of dark matter %spinor particles with arbitrary masses and chemical potentials. 
%The SDM fields may fluctuate with the temperature and is established around the equilibrium state at some weak background scale.    
%We have considered an effective description of a more complete model in which the SDM field is established at some background value. 
%The thermal fluctuation of the SDM field with the temperature (\ref{e20}) in the order of 3/2 depends crucially on the EW model parameters $\lambda$, $\kappa_D$ and $\phi_0$ that %can be completed from the LHC data.
%The $\chi$ DM mass $m$ is free parameter and is restricted within the relation (\ref{e21}). 
%The Higgs portal parameter $\kappa_D$ is in correlation with the temperature: the trivial result for $\tilde\varphi = 1$ will be at $T =0$ and $\kappa_D = 0$. 
The SDM associated with the condensate may be difficult to detect as it has the vacuum quantum numbers. The importance of the role concerning the SDM can be seeing when the stability of the BS and the scale invariance breaking are simultaneously disappear if the SDM fields are removing from the vacuum. This means the stability of the BS and the scale symmetry breaking are strongly correlated to each other via the SDM sector.
We have also considered an effective description of a more complete model with the (thermo)dynamic potential of the BS at finite temperature for an arbitrary number of dark matter spinor particles with arbitrary masses and chemical potentials. 
The thermal fluctuation of the SDM field with the temperature (\ref{e20}) depends crucially on the model parameters $\lambda$, $\kappa_D$ and $\phi_0$ that can be completed from the LHC data.
%The $\chi$ DM mass $m$ is free parameter and is restricted within the relation (\ref{e21}). 
%The Higgs portal parameter $\kappa_D$ is in correlation with the temperature: the trivial result for $\tilde\varphi = 1$ will be at $T =0$ and $\kappa_D = 0$. 
The observation for the BS traces  may extend the spectrum of research reconsidering a few tens of years exploration to discover the scalar hidden sector as well as the  scalar boson stars,  composed of the fields from this sector as possible candidates to DM. 
One can admit, if the SDM gravitationally clustered into cosmologically macroscopic bound states, this displays some universality with the SM sector. This can be seeing through the decays of the SDM because of instability of the BS as well as existence of the Bose-Einstein condensate.
Finally, we point out that there is now stronger motivation than ever to study the vicinity of the CP in order to understand the BS formation based on deep understanding of a rich and profound nature of interactions between the SDM fields with the GR at finite $T$.

\end{document}